# Title: Room-temperature chirality switching in a helimagnetic thin film

**Authors:** Hidetoshi Masuda[1*], Takeshi Seki[1*], Yoichi Nii[1,2,3], Hiroto Masuda[1], Koki Takanashi[1,4†], Yoshinori Onose[1*]

**Affiliations:**

[1]Institute for Materials Research, Tohoku University; Aoba-ku, Sendai, 980-8577, Japan.

[2]PRESTO, Japan Science and Technology Agency; Kawaguchi, Saitama, 332-0012, Japan.

[3]Organization for Advanced Studies, Tohoku University; Aoba-ku, Sendai, 980-8577, Japan.

[4]Center for Science and Innovation in Spintronics, Tohoku University; Aoba-ku, Sendai, 980-8577, Japan.

[*]Corresponding authors. Email: hidetoshi.masuda.c8@tohoku.ac.jp (Hidetoshi M.); takeshi.seki@tohoku.ac.jp (T. S.); yoshinori.onose.b4@tohoku.ac.jp (Y. O.)

[†]Present address: Advanced Science Research Center, Japan Atomic Energy Agency, Tokai, Ibaraki, 319-1195, Japan

**Abstract:** Helimagnetic structures, in which the magnetic moments are spirally ordered, host an internal degree of freedom called chirality (or helicity) corresponding to the handedness of the helix. The chirality seems quite robust against disturbances and is therefore promising for next-generation magnetic memory. While the chirality control was recently achieved by the magnetic field sweep with the application of an electric current at low temperature in a conducting helimagnet, problems such as low working temperature and cumbersome control sequence have to be solved in practical applications. Another issue is the thin film fabrication that enables the development of spintronic devices based on helimagnets. Here we show chirality switching by electric current pulses at room temperature in a thin-film $MnAu_2$ helimagnetic conductor. The result demonstrates the feasibility of helimagnet-based spintronics that can overcome all the above problems.

**Introduction:**

In solids, an internal degree of freedom emerges upon a phase transition involving symmetry breaking (Fig. 1). For example, a ferroelectric phase transition breaks the space inversion symmetry and induces spontaneous electric polarization. The two states with positive and negative polarizations are completely degenerate in the absence of electric fields, and therefore the polarization can be viewed as an internal degree of freedom. The application of electric fields can switch the sign of electric polarization. Such a controllable internal degree of freedom is useful for memorizing information and is therefore applicable to memory storage devices. Indeed, random-access memories based on ferroelectrics have been fabricated and are commercially available (*1*, *2*). A more important example is a ferromagnet. In the ferromagnetic state, the time-reversal symmetry is broken, and the magnetization is the internal degree of freedom, which can be controlled by a magnetic field. Hard disk drives utilize ferromagnets, and magnetic random-access memory (MRAM) has also been developed (*3*, *4*). One of the major obstacles for high-density MRAM is stray fields. As the bit scale is decreased, the magnetizations of separated ferromagnets do not work as independent degrees of freedom owing to the entanglement caused by stray fields. In order to resolve this issue, spintronics based on antiferromagnets is currently attracting considerable attention (*5-8*). A helical magnet (*9*) is one form of antiferromagnet that has unique characteristics: Mirror symmetry is broken, and the chirality works as an internal degree of freedom unless the crystal structure is noncentrosymmetric. The chirality does not couple to the magnetic field and is invariant under any translation and rotation. In order to reverse it, one has to first straighten the spin direction and wind it reversely. In other words, the helimagnetic memory is topologically protected in that way and should be stable even in a very small device. Therefore, this seems to be a desirable degree of freedom for next-generation magnetic storage. Nevertheless, the chirality in conducting helimagnets that are compatible with spintronic devices had been uncontrollable until recently, whereas the chirality in insulating helimagnets is known to be controllable with an electric field (*10*, *11*).

Recently Jiang *et al.* (*12*) showed that the degeneracy relevant to the chirality is lifted by the simultaneous application of magnetic fields and electric currents owing to the spin transfer torque (*13-15*), which is a totally different mechanism from the insulating case. The favored chirality depends on whether the electric current is parallel or antiparallel to the magnetic field. They demonstrated chirality control at temperatures around 50 K in a microfabricated single crystal piece of MnP helimagnetic conductor. To effectively control the chirality by utilizing the large susceptibility in the vicinity of the phase transition point, a complex sequence was adopted: first they applied a large electric current in the higher-field magnetic phase (fan magnetic phase in the case of

MnP) and decreased the magnetic field traversing the helical transition field. While the demonstration of chirality control is clear, their result cannot be directly applied to a practical device because of the low working temperature and the complex sequence of chirality control. In addition, a thin film form of helimagnet is needed to fabricate a practical device. In this work, we show room-temperature chirality control in a thin-film MnAu$_2$ helimagnetic by solely applying an electric current pulse in a magnetic field.

**Results:**

Properties of MnAu$_2$ thin film sample

MnAu$_2$ crystallizes in a centrosymmetric tetragonal crystal structure with the space group *I4/mmm* (*16*). The Mn magnetic moments show a helical magnetic order with a helical plane perpendicular to the propagation vector $\boldsymbol{q}$ = (0, 0, ~0.28) in the reciprocal lattice unit, corresponding to the helical pitch of ~3.1 nm (*17*). The transition temperature is reported to be as high as $T_c$ ~ 360 K (*18*). We prepared single-crystal films of MnAu$_2$ with a thickness of 100 nm on hexagonal ScMgAlO$_4$ (10−10) substrates. X-ray diffraction measurements revealed that the MnAu$_2$ thin film is stacked along the [110] direction so that the helical propagation vector is parallel to the thin film (Fig. 2A, see Materials and Methods and Fig. S1 for more detail). The magnetic susceptibility *M/H* along the [001] direction shows a clear kink at 335 K (Fig. 2B). The resistivity $\rho$ shows metallic temperature dependence with a cusp-like anomaly also around 335 K, as shown in Fig. 2C. The anomalies can be ascribed to the helical transition temperature (*18-20*), which is slightly lower than the reported value presumably because of epitaxial strain. Similar kinks appear in the magnetic field dependences of magnetization and resistivity. Fig. 3A shows the magnetization and magnetoresistance curves for *H* ∥ [001] at 300 K, in which clear kinks are observed at ~1.5 T. Above the kink field, the magnetization is saturated, which indicates that they are caused by the transition from the helical phase to the induced-ferromagnetic (FM) phase (*18*, *20*). The transition field increases as the temperature is lowered, as shown in Fig. 2D.

Chirality control by magnetic field sweep

We first demonstrate chirality control at room temperature by means of a magnetic field sweep with the application of an electric current, similarly to Jiang *et al* (*12*). A magnetic field $H_0$ = ±3 T and a dc electric current $j_0$ were first applied along the helical propagation vector. The magnetic field and the electric current were parallel ($H_0 > 0$, $j_0 > 0$ or $H_0 < 0$, $j_0 < 0$) or antiparallel ($H_0 > 0$, $j_0 < 0$ or $H_0 < 0$, $j_0 > 0$). Then, we swept the magnetic field to 0 T and turned off the electric current. After this control procedure, we read out the controlled chirality utilizing the nonreciprocal electronic transport (NET) (*21-23*), which is a field-asymmetric component of the 2nd-harmonic resistivity $\rho^{2\omega}_{asym}(H) = [\rho^{2\omega}(+H) - \rho^{2\omega}(-H)]/2$ under an ac electric current (*24-27*). Note that $\rho^{2\omega}$ has the same unit as the ordinary resistivity and is proportional to the applied current (See Fig. S2). The NET shows up only when the inversion and time-reversal symmetries are simultaneously broken and reverses its sign upon a space-inversion or time-reversal operation. Since the chirality is reversed upon the space-inversion operation, the sign of NET probes the chirality (*12, 21, 24-27*). In order to obtain $\rho^{2\omega}_{asym}(H)$, we measured $\rho^{2\omega}(H)$ while increasing $H$ from 0 T to 3 T and that also while decreasing $H$ from 0 T to -3 T and calculated the difference. Figure 3B shows the magnetic field dependence of the NET signal $\rho^{2\omega}_{asym}(H)$ after the field sweep chirality control with $H_0 = \pm 3$ T and $j_0 = 0, \pm 8.0 \times 10^9$ A/m². While $\rho^{2\omega}_{asym}(H)$ was almost negligible for $j_0 = 0$, finite $\rho^{2\omega}_{asym}(H)$ was observed for the other data. The magnitude steeply increases as the field magnitude is increased from 0 T. It shows a maximum around 0.5 T and almost vanishes above the ferromagnetic transition field. We confirm the magnetic field angle dependence of $\rho^{2\omega}_{asym}(H)$ is consistent with the chiral symmetry (see Figs. S2 and S3 for more detail). Importantly, the sign of the NET signal depends on whether $H_0$ and $j_0$ are parallel or antiparallel, confirming that the chirality was controlled successfully. The inset for Fig. 3B shows the $j_0$ dependence of the NET signal. The NET signal monotonically increases with $j_0$ and saturates around $j_0 = 6.0 \times 10^9$ A/m², suggesting that the chirality is controlled in nearly the full volume fraction.

Chirality switching by electric current pulses
Then, we performed the chirality control by the application of electrical current pulses. A theory in the literature (*28*) showed that the chirality can be controlled by the application of an electrical current pulse under a small magnetic field in the helimagnetic state when the magnitude of the electric current is much larger. The simple switching largely increases the availability of the chiral degree of freedom in spintronic devices. To experimentally demonstrate the chirality switching, we first swept the magnetic field from the high field to zero without an electric current so that the two chiral domains are almost equally distributed (see Fig. 3B), and then applied positive and negative electrical pulses with a duration of 1 ms alternately every 15 minutes at 0.5 T while measuring the second-harmonic resistivity. Figure 4A shows the time dependence of the second-harmonic resistivity change $\Delta \rho^{2\omega}$ after the zero current field sweep. When the pulse current $j_p$ was larger than $15 \times 10^9$ A/m², a discontinuous change of $\Delta\rho^{2\omega}$ appeared. The magnitude of the discontinuous change increases as the current is increased, and the negative pulse reversed $\Delta\rho^{2\omega}$. We observed the alternating change of $\Delta\rho^{2\omega}$ several times. To confirm that such a discontinuous change of the second-harmonic resistivity corresponds to the chiral domain change, we measured the field dependence of the second-harmonic resistivity after the application of an electric current pulse at 0.5 T. For this experiment, we also performed a zero-current magnetic sweep from 3 T to 0 T before the pulse application. After the pulse application, we decreased the magnetic field to -0.5 T and restored it to +0.5 T while measuring the second-harmonic resistivity. The measured second-harmonic resistivity data are shown in Fig. 4B. In this figure, the difference from the zero-field value is plotted just for clarity. The asymmetric field dependence is clearly observed, and its sign depends on that of the current pulse. These results show that the discontinuous change of the second-harmonic resistivity is certainly caused by the chiral domain change.

Finally we discuss the $j_p$ dependence of $\rho^{2\omega}_{asym}$ estimated from the magnetic field dependence after the pulse application (Fig. 4C). $\rho^{2\omega}_{asym}$ sharply increases around $1.4\times10^{10}$ A/m$^2$ and saturates around $1.9\times10^{10}$ A/m$^2$. For comparison, we reproduce the electric current dependence of the field sweep control. A much larger electric current is needed for the switching but the magnitude of the controlled $\rho^{2\omega}_{asym}$ is comparable with the sweeping case, suggesting that the controlled volume fraction is nearly full also for the pulse case. Thus, chiral domain switching is achieved for this thin film sample. Nevertheless, it should be noted that the heating effect seemed to assist the switching phenomenon. The sample resistance estimated from the voltage during the pulse application is larger than that at 300 K, indicating heating of the sample (see Fig. S4). The estimated sample temperature $T_{sam}$ is plotted against $j_p$. It gradually increases and exceeds the transition temperature 335 K around $2\times10^{10}$ A/m$^2$. When the nominal experimental temperature is decreased to 290 K the $j_p$ dependence is also shifted, which suggests that the heating effect certainly contributes to the chiral switching phenomenon (see Fig. S5). The critical current of chirality domain control is expected to decrease with increasing temperature toward $T_c$ = 335 K. Presumably, it becomes low enough around 315 K for the chirality switching, and finite $\rho^{2\omega}_{asym}$ is observed above $1.4\times10^{10}$ A/m$^2$. Traversing the helical transition temperature is not mandatory in this phenomenon.

**Discussion:**
In summary, we have demonstrated chirality switching at room temperature in a thin-film MnAu$_2$ helimagnet. The chirality degree of freedom is quite robust against magnetic disturbances and is free from the stray field problem. The threshold current density of $j_p \sim 1.4\times10^{10}$ A/m$^2$ for the chirality switching is relatively low compared to conventional antiferromagnets (*29-33*). In addition, the thin film form of the sample enables us to utilize the interface-based functionality. Although chirality control is performed in a magnetic field at present, zero-field control should be achievable in a helimagnet/ferromagnet hybrid device (*28*). Moreover, in chiral systems, there emerge novel spin responses denoted as "chirality-induced spin selectivity CISS" (*34*). The magnetic chiral system seems to also show similar properties (*35*). By utilizing the CISS response, we expect we will be able to more sensitively probe the chirality, which may enable us to satisfy the technological requirement regarding the reading out the memory. Thus, the present result clearly shows the potential of helimagnet-based spintronics.

**Materials and Methods:**

Sample fabrication

Epitaxial films of MnAu$_2$ with a thickness of 100 nm were deposited on ScMgAlO$_4$ (10−10) substrates by magnetron sputtering from Mn and Au targets at 400 °C. The film was then in-situ covered by a Ta cap layer with a thickness of 2 nm and annealed at 600 °C for 1 hour. The X-ray diffraction (XRD) results indicate the epitaxial growth of MnAu$_2$ (110) on the ScMgAlO$_4$ (10−10) substrate, where the MnAu$_2$ [001] and the ScMgAlO$_4$ [0001] directions are parallel to each other (Fig. S1).

Magnetization measurement

The magnetization measurements were performed using a Magnetic Property Measurement System (Quantum Design).

First- and second-harmonic resistivity measurement

For the resistivity measurements, the thin film sample was patterned into Hall bar devices by photolithography and Ar plasma etching. The direction of the electric current was parallel to the MnAu$_2$ [001] direction. The width of the Hall bars was 10 μm, and the distance between two voltage electrodes for the resistivity measurement was 25 μm. Electrical contacts were made by photolithography and electron beam evaporation of Ti (10 nm) / Au (100 nm). The resistivity measurements were performed in a superconducting magnet and a Physical Property Measurement System (Quantum Design). 1st- and 2nd-harmonic ac resistivities were measured by the lock-in technique with an electric current frequency of 11.15 Hz.

**References:**

1. H. Ishiwara, M. Okuyama, Y. Arimoto, *Ferroelectric Random Access Memories Fundamentals and Applications* (Springer, Berlin, Heidelberg, 2004)
2. H. Ishiwara, Ferroelectric random access memories. *J. Nanosci. Nanotechnol.* **12**, 7619-7627 (2012).
3. B. Dieny, R. B. Goldfarb, K. -J. Lee, *Introduction to Magnetic Random-Access Memory* (Wiley-IEEE Press, 2016).
4. S. Bhatti, R. Sbiaa, A. Hirohata, H. Ohno, S. Fukami, S. N. Piramanayagam, Spintronics based random access memory: a review. *Mater. Today* **20**, 530-548 (2017).
5. A. H. MacDonald, M. Tsoi, Antiferromagnetic metal spintronics. *Philos. Trans. R. Soc. A* **369**, 3098-3114 (2011).
6. E. V. Gomonay, V. M. Loktev, Spintronics of antiferromagnetic systems (Review Article). *Low Temp. Phys.* **40**, 17-35 (2014).
7. T. Jungwirth, X. Marti, P. Wadley, J. Wunderlich, Antiferromagnetic spintronics. *Nat. Nanotechnol.* **11**, 231-241 (2016).
8. V. Baltz, A. Manchon, M. Tsoi, T. Moriyama, T. Ono, Y. Tserkovnyak, Antiferromagnetic spintronics. *Rev. Mod. Phys.* **90**, 015005 (2018).
9. A. Yoshimori, A new type of antiferromagnetic structure in the Rutile type crystal. *J. Phys. Soc. Jpn.* **14**, 807-821 (1959).
10. S. W. Cheong, M. Mostovoy, Multiferroics: a magnetic twist for ferroelectricity. *Nat. Mater.* **6**, 13-20 (2007).
11. Y. Tokura, S. Seki, Multiferroics with spiral spin orders. *Adv. Mater.* **22**, 1554-1565 (2010).
12. N. Jiang, Y. Nii, H. Arisawa, E. Saitoh, Y. Onose, Electric current control of spin helicity in an itinerant helimagnet. *Nat. Commun.* **11**, 1601 (2020).
13. G. Tatara, H. Kohno, J. Shibata, Microscopic approach to current-driven domain wall dynamics. *Phys. Rep.* **468**, 213-301 (2008).
14. P. M. Haney, R. A. Duine, A. S. Núñez, A. H. MacDonald, Current-induced torques in magnetic metals: beyond spin-transfer. *J. Magn. Magn. Mater.* **320**, 1300-1311 (2008).
15. O. Wessely, B. Skubic, L. Nordström, Spin-transfer torque in helical spin- density waves. *Phys. Rev. B* **79**, 104433 (2009).
16. O. E. Hall, J. Royan, The structure of Au$_2$Mn. *Acta Crystallographica* **12**, 607-608 (1959).
17. A. Herpin, P. Meriel, Étude de l'antiferromagnétisme helicoïdal de MnAu$_2$ par diffraction de neutrons. *J. Phys. Le Radium* **22**, 337-348 (1961).

18. A. J. P. Meyer, P. Taglang, Propriétés magnétiques, antiferromagnétisme et ferromagnétisme de MnAu$_2$. *J. Phys. Le Radium* **17**, 457-465 (1956).
19. J. H. Smith, R. Street, Resistivity and magnetoresistance of Au$_2$Mn. *Proc. Phys. Soc. Sect. B* **70**, 1089-1092 (1957).
20. H. Samata, N. Sekiguchi, A. Sawabe, Y. Nagata, T. Uchida, M. D. Lan, Giant magnetoresistance observed in metamagnetic bulk-MnAu$_2$. *J. Phys. Chem. Solids* **59**, 377-383 (1998).
21. G. L. J. A. Rikken, J. Fölling, P. Wyder, Electrical magnetochiral anisotropy. *Phys. Rev. Lett.* **87**, 236602 (2001).
22. G. L. J. A. Rikken, P. Wyder, Magnetoelectric anisotropy in diffusive transport. *Phys. Rev. Lett.* **94**, 016601 (2005).
23. Y. Tokura, N. Nagaosa, Nonreciprocal responses from non-centrosymmetric quantum materials. *Nat. Commun.* **9**, 3740 (2018).
24. F. Pop, P. Auban-Senzier, E. Canadell, G. L. J. A. Rikken, N. Avarvari, Electrical magnetochiral anisotropy in a bulk chiral molecular conductor. *Nat. Commun.* **5**, 3757 (2014).
25. T. Yokouchi, N. Kanazawa, A. Kikkawa, D. Morikawa, K. Shibata, T. Arima, Y. Taguchi, F. Kagawa, Y. Tokura Electrical magnetochiral effect induced by chiral spin fluctuations. *Nat. Commun.* **8**, 866 (2017).
26. R. Aoki, Y. Kousaka, Y. Togawa, Anomalous nonreciprocal electrical transport on chiral magnetic order. *Phys. Rev. Lett.* **122**, 57206 (2019).
27. G. L. J. A. Rikken, N. Avarvari, Strong electrical magnetochiral anisotropy in tellurium. *Phys. Rev. B* **99**, 245153 (2019).
28. J. I. Ohe, Y. Onose, Chirality control of the spin structure in monoaxial helimagnets by charge current. *Appl. Phys. Lett.* **118**, (2021).
29. P. Wadley, B. Howells, J. Železný, C. Andrews, V. Hills, R. P. Campion, V. Novák, K. Olejník, F. Maccherozzi, S. S. Dhesi, S. Y. Martin, T. Wagner, J. Wunderlich, F. Freimuth, Y. Mokrousov, J. Kuneš, J. S. Chauhan, M. J. Grzybowski, A. W. Rushforth, K. W. Edmonds, B. L. Gallagher, T. Jungwirth, Electrical switching of an antiferromagnet. *Science* **351**, 587-590 (2016).
30. S. Y. Bodnar, L. Šmejkal, I. Turek, T. Jungwirth, O. Gomonay, J. Sinova, A. A. Sapozhnik, H. -J. Elmers, M. Kläui, M. Jourdan, Writing and reading antiferromagnetic Mn$_2$Au by Néel spin-orbit torques and large anisotropic magnetoresistance. *Nat. Commun.* **9**, 348 (2018).
31. J. Godinho, H. Reichlová, D. Kriegner, V. Novák, K. Olejník, Z. Kašpar, Z. Šobáň, P. Wadley, R. P. Campion, R. M. Otxoa, P. E. Roy, J. Železný, T. Jungwirth, J.

Wunderlich, Electrically induced and detected Néel vector reversal in a collinear antiferromagnet. *Nat. Commun.* **9**, 4686 (2018).

32. T. Moriyama, K. Oda, T. Ohkochi, M. Kimata, T. Ono, Spin torque control of antiferromagnetic moments in NiO. *Sci. Rep.* **8**, 14167 (2018).
33. H. Tsai, T. Higo, K. Kondou, T. Nomoto, A. Sakai, A. Kobayashi, T. Nakano, K. Yakushiji, R. Arita, S. Miwa, Y. Otani, S. Nakatsuji, Electrical manipulation of a topological antiferromagnetic state. *Nature* **580**, 608-613 (2020).
34. R. Naaman, D. H. Waldeck, Chiral-induced spin selectivity Effect. J. Phys. Chem. Lett. **3**, 2178-2187 (2012).
35. H. Watanabe, K. Hoshi, J. Ohe, Chirality-induced spin current through spiral magnets. Phys. Rev. B **94**, 125143 (2016).

**Acknowledgments:** The authors thank T. Sasaki for her help in doing the film deposition. The film deposition and device fabrication were carried out at the Cooperative Research and Development Center for Advanced Materials, IMR, Tohoku University.

**Funding:** This work was partially supported by JSPS KAKENHI grant numbers JP20H00299, JP20K03828, JP21H01036, and JP22H04461, JST PRESTO grant number JPMJPR19L6, and the Mitsubishi Foundation.

**Author Contributions:** T. S. grew the film and performed x-ray diffraction under the support by K. T. Hidetoshi M. performed the device fabrication with the use of photolithography and Ar plasm etching, 1st and 2nd harmonic resistivity measurements, and magnetization measurements with supports from T. S., Y. N., and Hiroto M. Y. O. conceived the project. Hidetoshi M. and Y. O. wrote the draft with input from T. S., Y. N., Hiroto M., and K. T.

**Competing Interests:** The authors declare no competing interests.

**Supplementary Materials**
Figs. S1 – S5

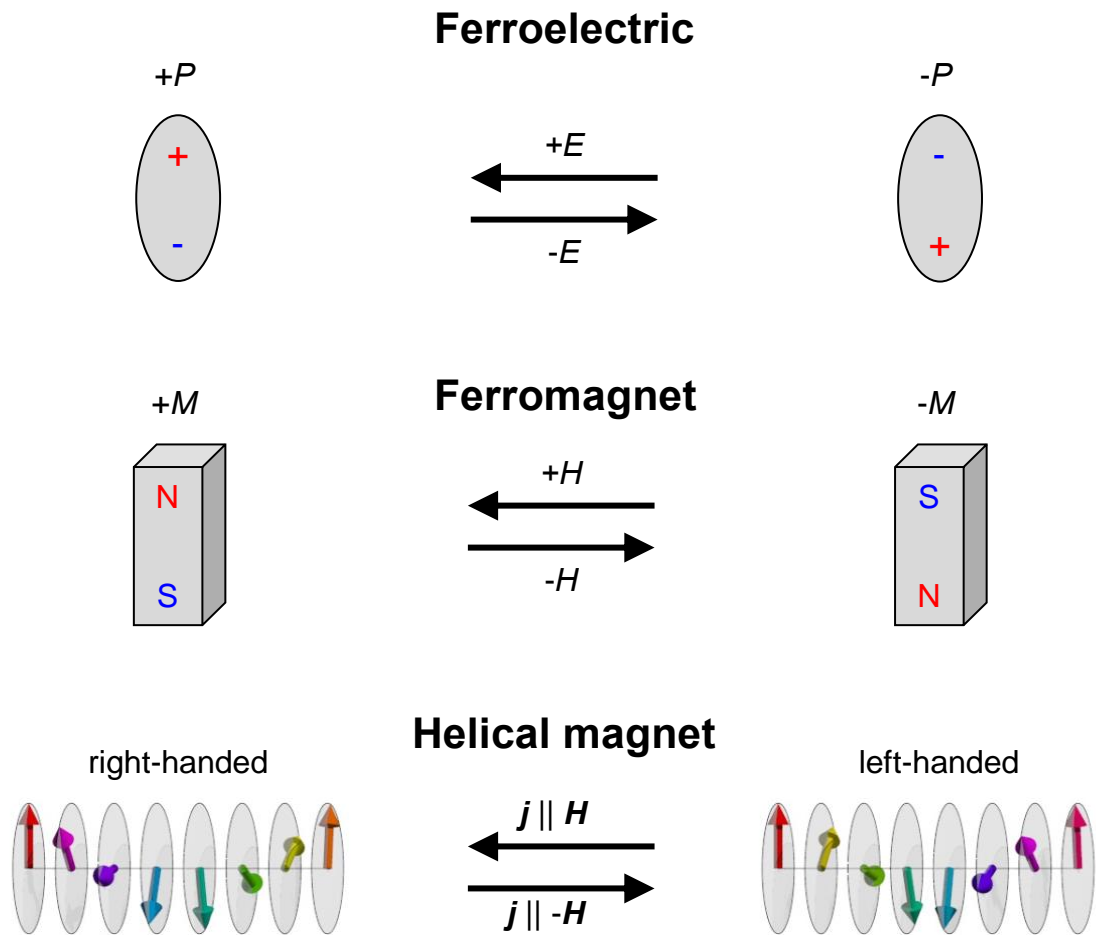

**Fig. 1. Degrees of freedom and their control with external fields in solids.**
Schematic illustrations of electric field control of polarization in a ferroelectric (top), magnetic field control of magnetization in a ferromagnet (middle), and chirality control with an electric current and a magnetic field in a helimagnet (bottom). In a helimagnet, the chirality depends on whether the electric current and the magnetic field are parallel or antiparallel.

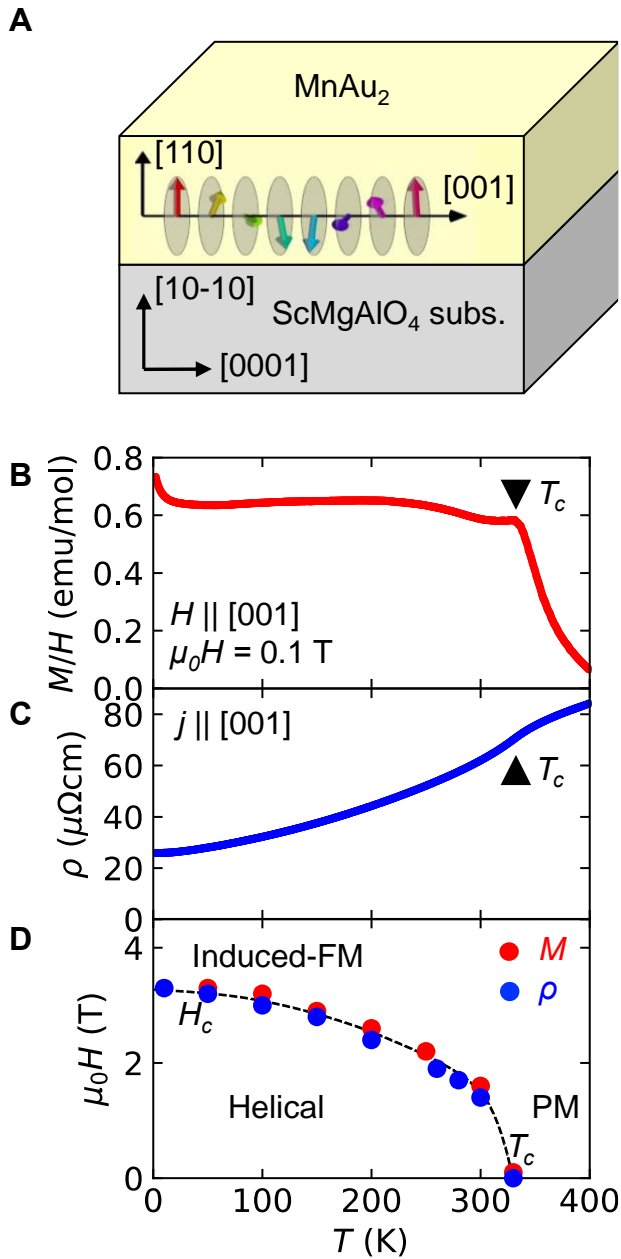

**Fig. 2. Properties of the MnAu₂ thin film sample.**
(A) Schematic illustration of the MnAu₂ thin film sample. MnAu₂ was epitaxially grown along the [110] direction with a thickness of 100 nm on a ScMgAlO₄ (10−10) substrate. The propagation vector of the helimagnetic structure is parallel to the [001] direction of MnAu₂ in the sample plane. The Ta cap layer (2 nm) is not shown for clarity.
(B) Temperature $T$ dependence of the magnetic susceptibility $M/H$, which is obtained by the magnetization $M$ divided by the magnetic field $H$. The magnetic field is applied along the [001] direction.

**(C)** $T$ dependence of the resistivity $\rho$. The electric current is applied along the [001] direction.

**(D)** Magnetic phase diagram in the $H$-$T$ plane for the MnAu$_2$ thin sample. The magnetic transition points are obtained by the magnetization and resistivity measurements. PM and induced-FM denote the paramagnetic and field-induced ferromagnetic states, respectively. $H_c$ denotes the transition field from the helical phase to the induced-FM phase. The dashed line is merely a guide for the eyes.

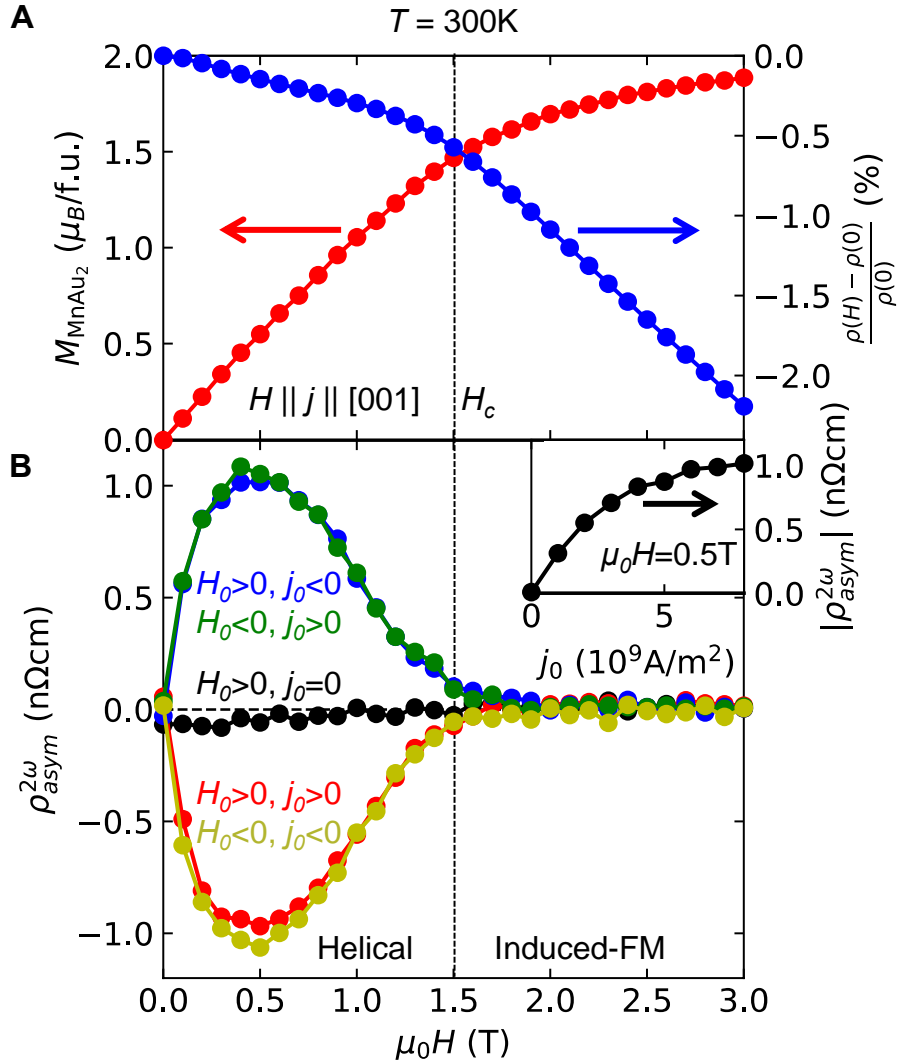

**Fig. 3. Chirality control by magnetic field sweep with application of electric current.**

(A) Magnetic field $H$ dependence of the magnetization $M_{MnAu_2}$ and the magnetoresistance $[\rho(H)-\rho(0)]/\rho(0)$ at $T = 300$ K. The linear diamagnetic contribution from the substrate is subtracted from the magnetization. The vertical dashed line denotes $H_c$.

(B) Magnetic field dependence of $\rho^{2\omega}_{asym}(H)$ at 300 K after the chirality control by the magnetic field sweep from the magnetic field $H_0$ ($H_0 = \pm 3$ T) with the application of electric current $j_0 = 0, \pm 8.0 \times 10^9$ A/m$^2$. $\rho^{2\omega}_{asym}(H)$ is the anti-symmetric part of the 2nd-harmonic electrical resistivity $\rho^{2\omega}(H)$, viz., $\rho^{2\omega}_{asym}(H) = [\rho^{2\omega}(+H) - \rho^{2\omega}(-H)]/2$. $\rho^{2\omega}(+H)$ and $\rho^{2\omega}(-H)$ were independently measured just after the chirality control. The ac electric current used for the $\rho^{2\omega}(H)$ measurement is $2.0 \times 10^9$ A/m$^2$. The inset shows the $j_0$ dependence of $|\rho^{2\omega}_{asym}(H)|$ at 0.5 T.

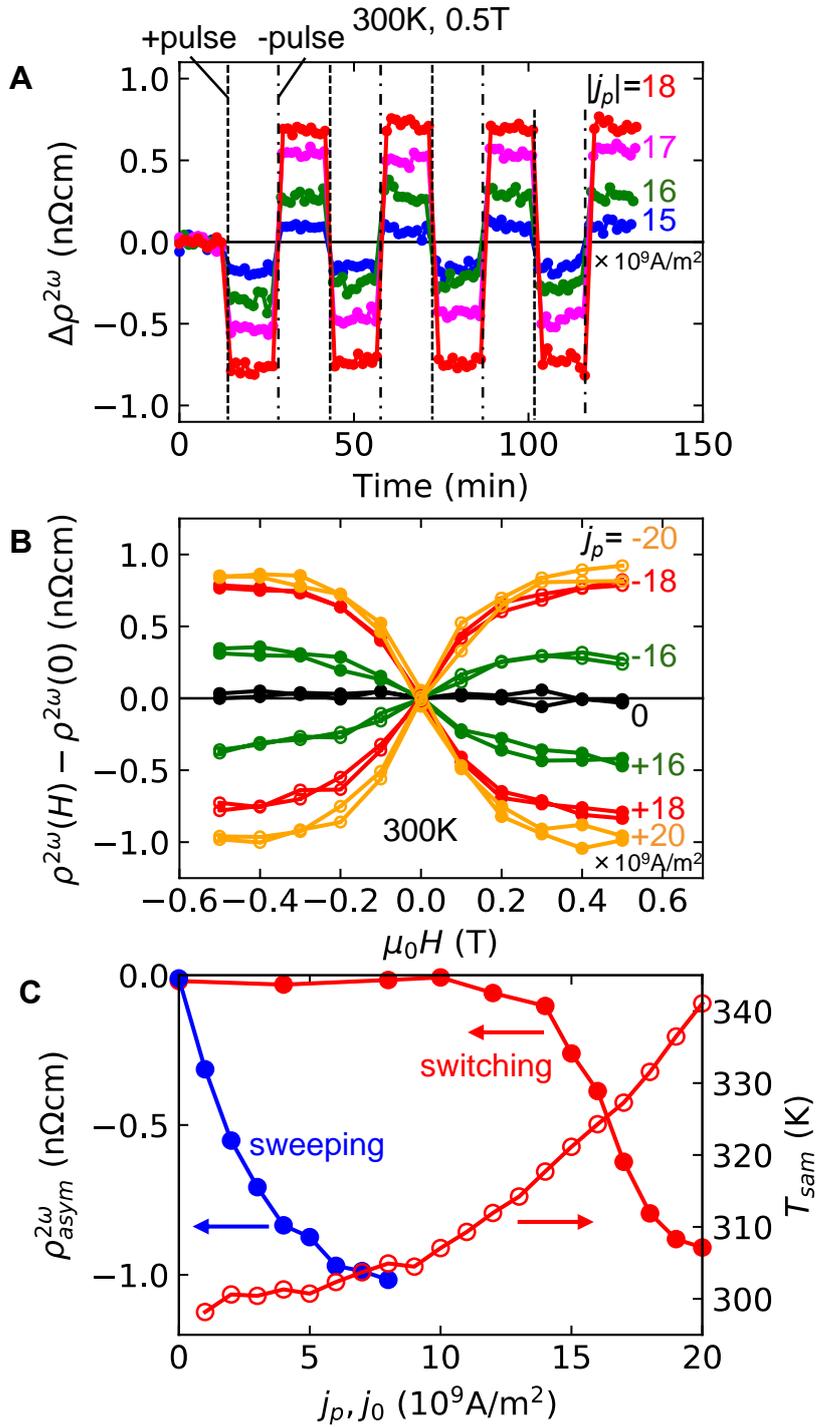

**Fig. 4. Chirality switching.**
(A) Temporal variation of second harmonic resistivity change $\Delta\rho^{2\omega}$ with the alternate application of positive and negative electric current pulses with a duration of 1 ms. Before the measurement, the magnetic field is swept from 3 T without electric current so that two different chirality domains are equally distributed. The magnitudes of the pulse currents are $|j_p|$ = 15 (blue), 16 (green), 17 (magenta), and 18(red) $\times 10^9$ A/m².

**(B)** Magnetic field dependence of $\rho^{2\omega}$ after application of the current pulses at 0.5 T. The pulse currents are $j_p$ = 0 (black), +16×10⁹ A/m² (green, filled), −16×10⁹ A/m² (green, open), +18×10⁹ A/m² (red, filled), −18×10⁹ A/m² (red, open), +20×10⁹ A/m² (orange, filled) and −20×10⁹ A/m² (orange, open). Before the measurement, the magnetic field is swept from 3 T to 0 T without an electric current, similarly to the case of Fig. 4A.

**(C)** $\rho^{2\omega}_{asym}$ at 0.5 T as a function of $j_p$ (red-filled circles). The $j_0$ dependence of $\rho^{2\omega}_{asym}$ in the case of field sweep control is reproduced from Fig. 3B for comparison (blue-filled circles). The red-open circles indicate the sample temperature $T_{sam}$ during the applied current pulse $j_p$ estimated from the sample resistance (see Fig. S4 for more detail).

Supplementary Materials for:

# Room-temperature chirality switching

# in a helimagnetic thin film

Hidetoshi Masuda*, Takeshi Seki*, Yoichi Nii, Hiroto Masuda, Koki Takanashi, Yoshinori Onose*

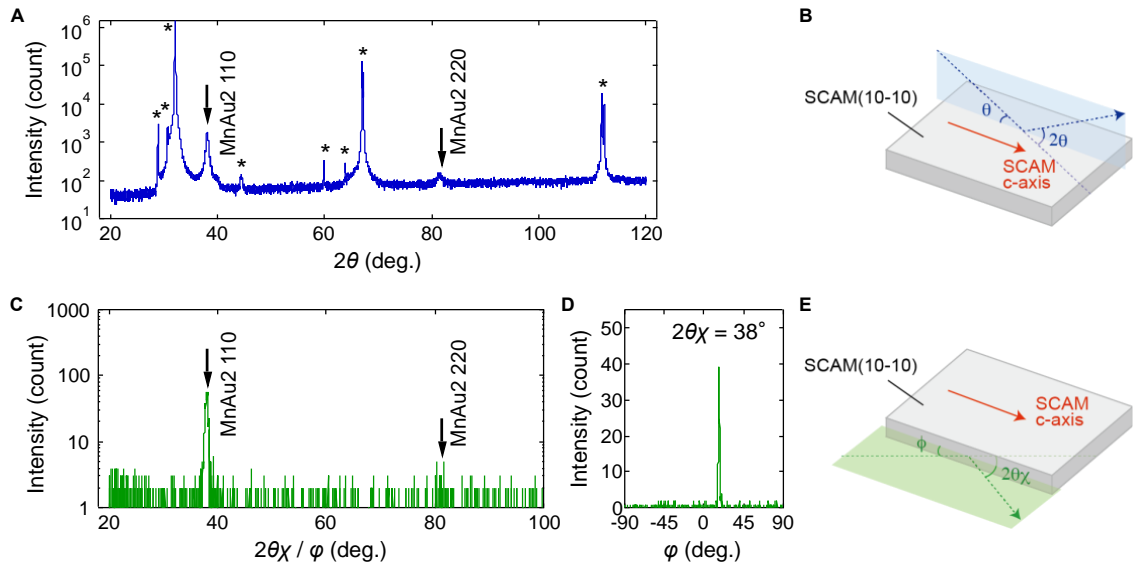

**Fig. S1. XRD profiles of the ScMgAlO$_4$ (10−10) / MnAu$_2$ (110) [100nm] / Ta[2nm] thin film sample.**

(**A**) Out-of-plane $\theta/2\theta$-scan. MnAu$_2$ (110) and (220) reflections indicate that the MnAu$_2$ (110) plane is grown on the ScMgAlO$_4$ (10−10) plane. Asterisks denote the peaks from the SCAM (ScMgAlO$_4$) substrate or sample stage. No extra reflections were observed.
(**B**) Schematic illustration of the out-of-plane $\theta/2\theta$-scan.
(**C**) In-plane $2\theta\chi/\varphi$-scan. The observation of MnAu$_2$ (110) and (220) reflections indicates that the MnAu$_2$ [1−10] direction is perpendicular to the ScMgAlO$_4$ [0001] direction (see Fig. S1E).
(**D**) In-plane $\varphi$-scan. A single sharp peak in the $\varphi$-scan range from −90° to 90° indicates the in-plane single-crystal order.
(**E**) Schematic illustration of the in-plane $2\theta\chi/\varphi$- and $\varphi$- scans.

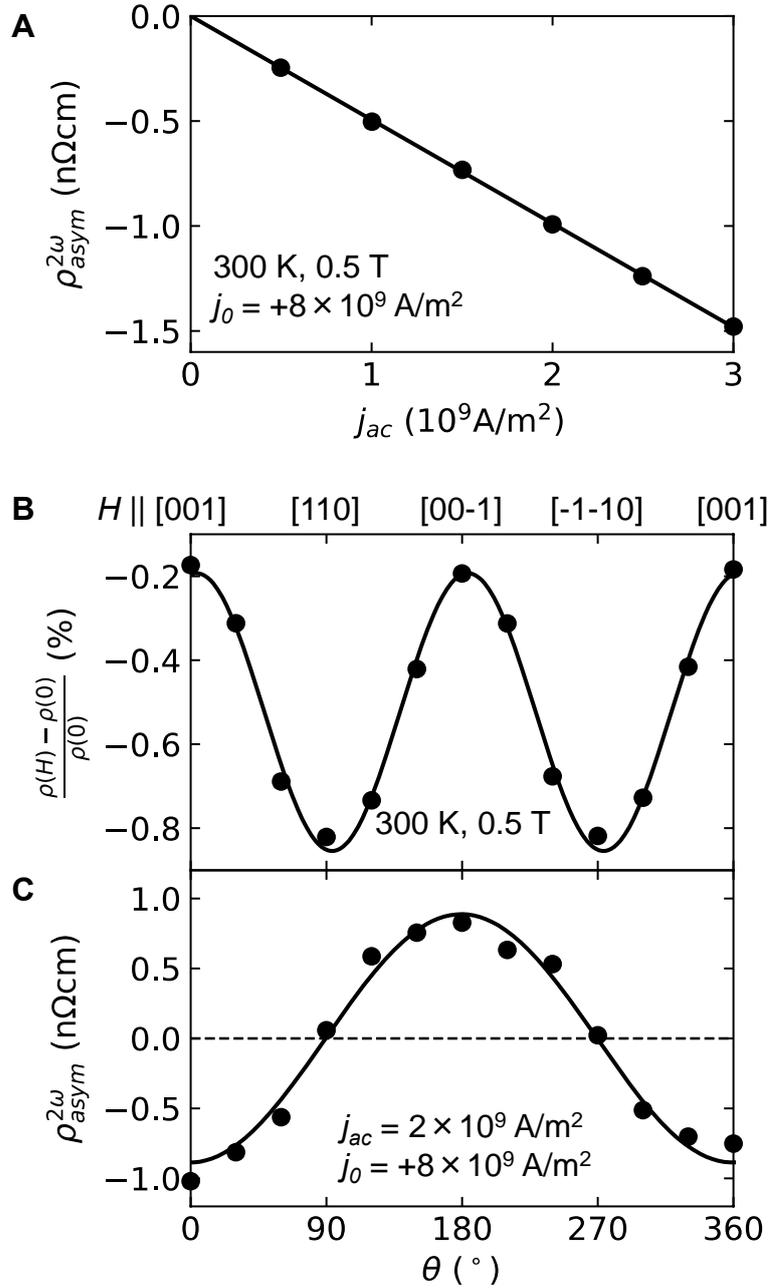

**Fig. S2. Properties of nonreciprocal electronic transport in MnAu$_2$.**
(**A**) The ac current $j_{ac}$ dependence of $\rho^{2\omega}_{asym}$ at 300 K and 0.5 T after the field sweep from $H_0 = +3$ T with $j_0 = +8.0\times10^9$ A/m$^2$. $\rho^{2\omega}_{asym}$ shows a linear $j_{ac}$ dependence, being consistent with the picture of nonreciprocal electric transport.
(**B**) Magnetoresistance $[\rho(H) - \rho(0)]/\rho(0)$ as a function of magnetic field angle $\theta$ at 300 K and 0.5 T. Here $\theta$ is the angle between the current and the magnetic field. The magnetic field is rotated within the (1−10) plane. The solid line is the result of fitting to $\cos2\theta$. The 180-degree rotation of the magnetic field does not alter the linear resistivity.

**(C)** $\theta$ dependence of $\rho^{2\omega}{}_{asym}$ at 300 K and 0.5 T after the field sweep from $H_0 = +3$ T with $j_0 = +8.0\times10^9$ A/m². $\rho^{2\omega}{}_{asym}$ shows $\cos\theta$ angle dependence, and the magnitude is maximum at $\theta = 0$ and 180 deg., being consistent with the chiral symmetry.

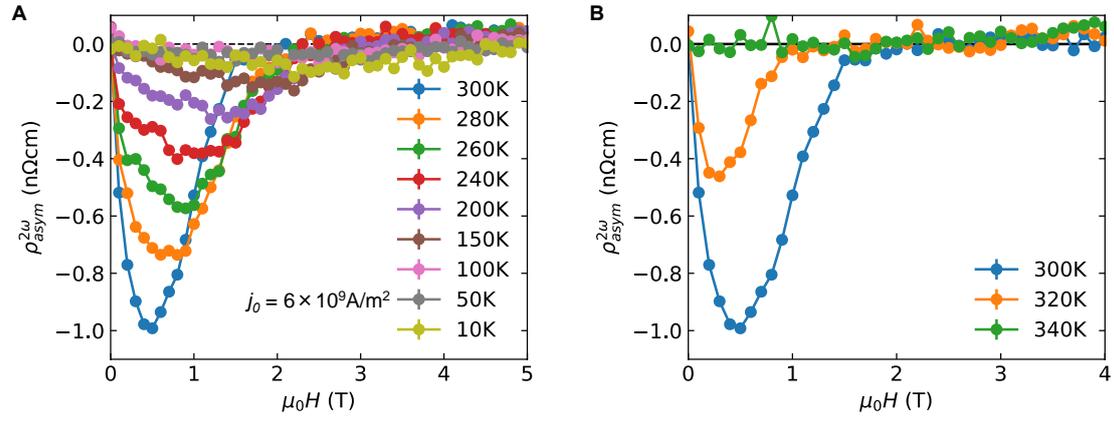

**Fig. S3. Field sweep chirality control at various temperatures.**
(**A, B**) Magnetic field dependence of $\rho^{2\omega}_{asym}(H)$ after the field sweep from $H_0 = +5$ T with $j_0 = +6.0\times10^9$ A/m$^2$ at various temperatures. The field sweep control and the measurement of $\rho^{2\omega}_{asym}(H)$ are performed at the same temperature.

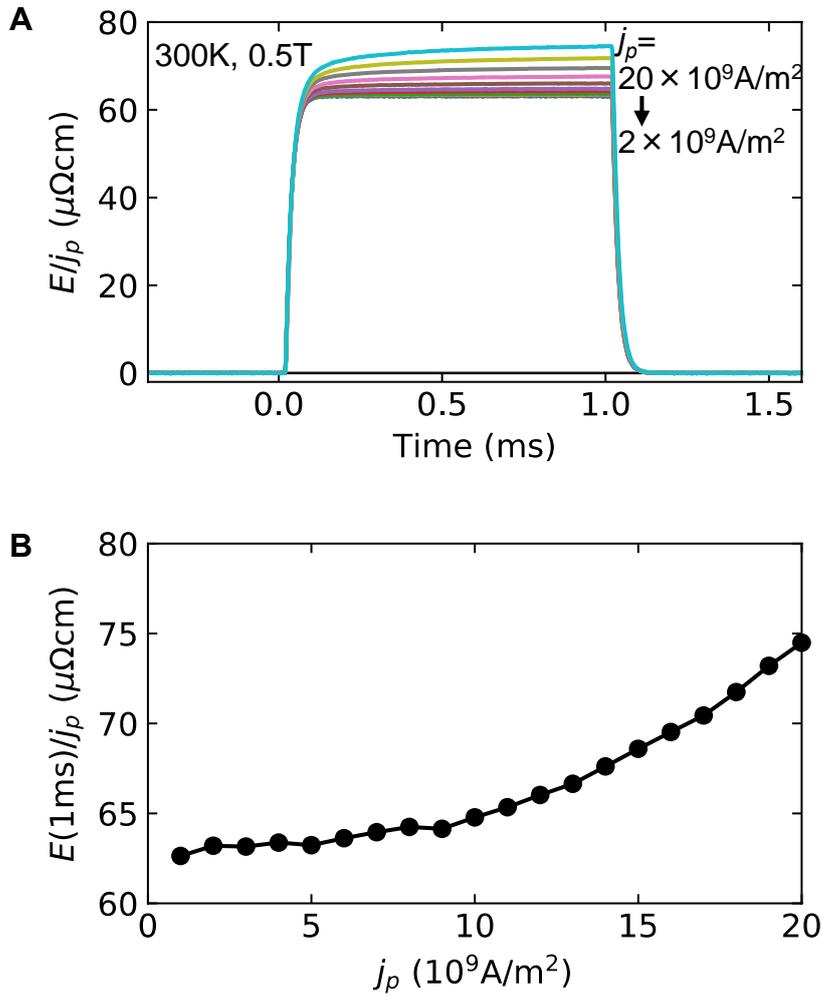

**Fig. S4. Estimation of sample heating by the application of electric current pulse.**
**(A)** Time dependence of the resistivity $E/j_p$ during the current pulse application with various magnitudes.
**(B)** $j_p$ dependence of resistivity $E/j_p$. In this figure, $E/j_p$ at 1 ms is adopted. We estimated the sample temperature during the pulse application from these data and the temperature dependence of resistivity shown in Fig. 2B.

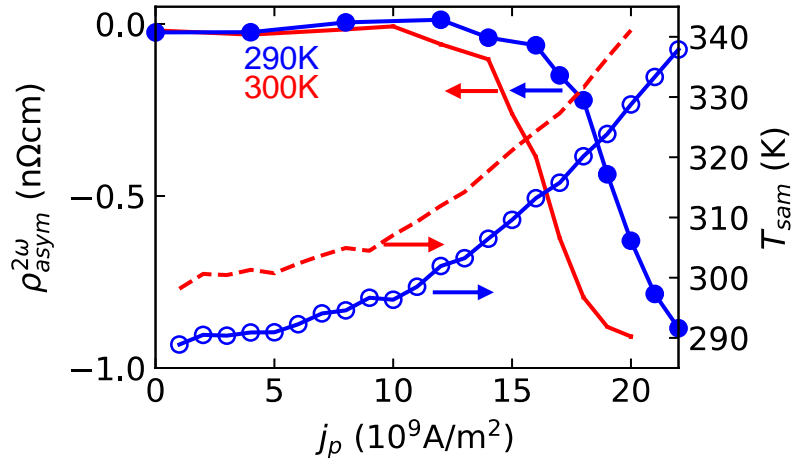

**Fig. S5. Comparison of instantaneous chirality switching results at 290 K and 300 K.** The $j_p$ dependence of $\rho^{2\omega}_{asym}$ and the sample temperature $T_{sam}$ during the applied current pulse at nominal experimental temperatures 290 K and 300 K. The experimental probe was controlled to the nominal temperature, but the sample was heated by the electric current pulse. When the nominal temperature is decreased to 290 K, the $j_p$ dependences are horizontally shifted, indicating that the sample temperature is important for the $j_p$ dependence.